\documentclass[reprint,10pt,showpacs,groupedaddress,superscriptaddress,floatfix,notitlepage]{revtex4-2}
\usepackage{amsmath,amsfonts,amssymb,graphics,graphicx,epsfig,color,times, braket}
\usepackage[utf8x]{inputenc}
\usepackage{color}
\usepackage{bbm, dsfont}
\usepackage{subfigure}
\usepackage{hyperref} 
\usepackage{mathrsfs}
\usepackage{verbatim} 
\usepackage{booktabs}
\makeatletter
\newcommand{\equalcontrib}{\frontmatter@footnote{These authors contributed equally to this work.}}
\renewcommand{\section}{%
  \@startsection{section}{1}{\z@}%
    {0.8cm \@plus1ex \@minus .2ex}%
    {1sp}%
    {\normalfont\small\bfseries\raggedright}%
}
\renewcommand{\subsection}{%
  \@startsection{subsection}{2}{\z@}%
    {0.8cm \@plus1ex \@minus .2ex}%
    {1sp}%
    {\normalfont\small\bfseries\raggedright}%
}
\makeatother
\begin{document}
\title{Active Approach to Fast Quantum State Preparation and Reset in Atomic Arrays}

\author{Ya-Tang~Yu\equalcontrib}
\email{yatan1018@gmail.com}
\affiliation{Institute of Atomic and Molecular Sciences, Academia Sinica, Taipei 10617, Taiwan}
\affiliation{Department of Physics and Astronomy, Purdue University, West Lafayette, Indiana 47907, USA}

\author{Wei-Hsuan~Chung\equalcontrib}
\email{b12202069@ntu.edu.tw}
\affiliation{Institute of Atomic and Molecular Sciences, Academia Sinica, Taipei 10617, Taiwan}
\affiliation{Department of Physics, National Taiwan University, Taipei 10617, Taiwan}

\author{Tian-Shu~Gou}
\affiliation{Institute of Atomic and Molecular Sciences, Academia Sinica, Taipei 10617, Taiwan}
\affiliation{Graduate Institute of Electronics Engineering, National Taiwan University, Taipei 10617, Taiwan}


\author{Kuan-Ting~Lin}
\affiliation{Trapped-Ion Quantum Computing Laboratory, Hon Hai Research Institute, Taipei 11492, Taiwan}



\author{H.~H.~Jen}
\affiliation{Institute of Atomic and Molecular Sciences, Academia Sinica, Taipei 10617, Taiwan}
\affiliation{Physics Division, National Center for Theoretical Sciences, Taipei 10617, Taiwan}
\affiliation{Department of Physics, National Central University, Taoyuan City 320317, Taiwan}

\date{\today}
\renewcommand{\r}{\mathbf{r}}
\newcommand{\f}{\mathbf{f}}
\renewcommand{\k}{\mathbf{k}}
\def\p{\mathbf{p}}
\def\q{\mathbf{q}}
\def\bea{\begin{eqnarray}}
\def\eea{\end{eqnarray}}
\def\ba{\begin{array}}
\def\ea{\end{array}}
\def\bdm{\begin{displaymath}}
\def\edm{\end{displaymath}}
\def\red{\color{red}}
\pacs{}

\begin{abstract}
Rapid preparation of a target quantum state and reset to the many-body ground state are essential for quantum information processing. Here, we investigate both processes in a waveguide-coupled emitter array in the low-excitation regime. By decomposing the effective non-Hermitian Hamiltonian into biorthogonal eigenmodes, we show that relaxation depends on the initial overlap with the more subradiant modes, providing a modal explanation of the Mpemba effect. Guided by this insight, we design a reset protocol in which a finite-duration local drive activelysuppresses the relatively slow-decaying components. For the parameters studied, this reduces the time to reach a fixed residual-excitation threshold by up to approximately \(25\%\) compared with free decay. To prepare a target Dicke state, we introduce a two-step protocol that initially leaves one emitter undriven before restoring the full drive that stabilizes the target. This local control redistributes spectral weight into the more subradiant modes, driving them closer to their target amplitudes, which reduces the preparation time by up to approximately \(45\%\) while consuming less driving energy than continuous driving. Finally, we show how directional waveguide coupling shifts collective modes toward a preferred boundary through the non-Hermitian skin effect, providing guidance for selecting the preferred site for local operations. We term this fundamental dependence of the relaxation dynamics on the system's chirality the ``non-Hermitian Skin-Mpemba effect.'' These results demonstrate how selective local control can accelerate state preparation and reset in open quantum systems.

\end{abstract}
\maketitle


The precise preparation of quantum states, ranging from single-qubit initialization to the generation of complex entangled states, is the starting point for nearly every task in quantum computing and quantum information processing \cite{Yuan2022, Zhang2021, Araujo2020, Buhrman2023, Zhang2022, Zhang2023, Zheng2022}. Although preparation protocols are often evaluated primarily by their fidelity, the time required to reach the target state is equally critical \cite{Bond2024, Bukov2019, Shao2016}. Minimizing this initialization time accelerates the overall computational process and maximizes the repetition rate of the quantum processor. Furthermore, state reset is just as essential as initialization \cite{Yuan2023, Geerlings2013, Wald2024, Egger2018}. After a sequence of operations, residual excitations and correlations must be completely removed before the next computational cycle can commence, making rapid and reliable reset a key requirement for scalable quantum technologies.


In waveguide-coupled systems, a highly effective strategy for preparing a target excitation is to inject a driving field whose temporal profile is the exact time reversal of the natural emission from that state \cite{Liao2018, Lin2022, Srinivasan2014}. In the ideal limit, achieving perfect state preparation requires capturing the complete time-reversed waveform, which formally extends over an infinite duration. In practice, the injected waveform must be truncated: increasing its duration improves the preparation fidelity, whereas shortening it necessarily introduces truncation error. Consequently, once a desired fidelity is specified, the required preparation time is largely fixed, leaving very limited freedom to further accelerate the protocol.

A route beyond this time--fidelity constraint is suggested by the Mpemba effect \cite{Zhang2026, Zhang2025, Carollo2021, Ares2025, Xu2026, Klich2019, Lin2023, Chatterjee2024}. This phenomenon demonstrates that relaxation is not determined solely by the initial distance from equilibrium: under suitable conditions, a state initially farther from equilibrium can actually relax faster than a closer one. Such counterintuitive behavior arises because different initial states have varying overlaps with the system's relaxation modes. Motivated by this observation, we propose an active strategy to accelerate the resetting process. Rather than relying on unassisted relaxation, we apply an external drive for a finite duration to a waveguide-coupled array of two-level emitters in the low-excitation regime.

By actively reshaping the collective excitation before allowing the system to relax freely, we deliberately steer the array toward configurations with weaker overlap with slowly decaying modes. Although this operation temporarily injects energy and may move the system farther from its target, the complete process finishes sooner than natural relaxation from the original state, which is termed as the Pontus Mpemba effect \cite{Nava2025}. Applying this framework, we also examine the inverse task of target-state preparation, showing that active control can reduce both the preparation time and the required driving energy in suitable regimes. Finally, we investigate how the directionality of the waveguide influences the spatial profile of the collective eigenmodes through the non-Hermitian skin effect, providing guidance for selecting preferred sites for local operations. We term this fundamental dependence of the relaxation dynamics on the system's chirality the ``non-Hermitian Skin-Mpemba effect.'' Because this phenomenon arises from generic non-Hermitian dynamics, it is expected to be a universal feature across diverse chiral platforms, serving as a highly valuable resource for designing practical, high-speed quantum control applications.


\begin{figure}[t]
    \centering
    \IfFileExists{schematic.png}{%
        \includegraphics[width=\columnwidth]{schematic.png}%
    }{%
        \IfFileExists{schematic_clean.pdf}{%
            \includegraphics[width=\columnwidth]{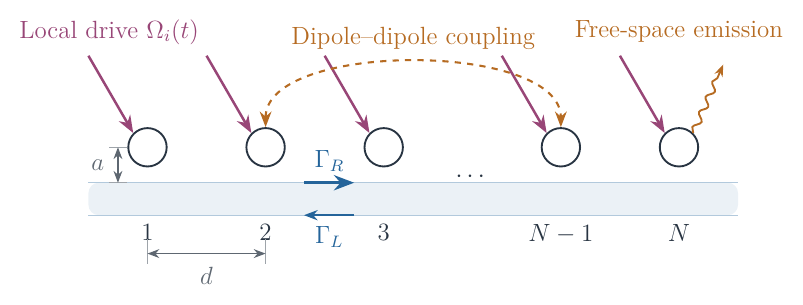}%
        }{%
            \fbox{\parbox[c][5.5cm][c]{0.95\columnwidth}{\centering
            Figure 1 placeholder\\[0.5em]
            Upload the figure as \texttt{schematic.pdf} or \texttt{schematic.png}.}}%
        }%
    }
    \caption{Schematic of the 1D atomic array in waveguide-QED setup. 
    An array of $N$ identical two-level atoms, with decay rate $\Gamma_{L(R)}$ to the left (right), is coupled to a one-dimensional chiral waveguide with interatomic separation $d$. The distance $a$ between atoms and waveguide can be controlled to adjust the directional coupling efficiencies $\beta$. Individual atoms can be coherently driven by external local fields with Rabi frequencies $\Omega_\mu(t)$}. 
    \label{fig:schematic}
\end{figure}






\section*{Results}

\subsection*{Mpemba Effect in a Waveguide-Coupled Atomic Array}
We consider a one-dimensional array of $N$ two-level emitters coupled to an infinite chiral waveguide (see Fig.~\ref{fig:schematic}). To account for imperfect coupling, we introduce a dimensionless parameter $\beta \in [0,1]$ \cite{Lodahl2017}. This effective emitter--waveguide coupling depends on the specific physical platform \cite{Sheremet2023, Mirhosseini2019, Arcari2014} and can be tuned by varying the spatial separation between the emitter array and the waveguide \cite{Sheremet2023, AsenjoGarcia2017}. The limit $\beta=1$ corresponds to ideal coupling into the guided mode, whereas $\beta=0$ represents an array completely decoupled from the waveguide, where the system evolves entirely via the remaining photon-mediated dipole-dipole interactions \cite{Gross1982,Lehmberg1970,Jen2025}.

The full open-system dynamics originate from a master equation derived under the Born--Markov and rotating-wave approximations \cite{Pichler2015, Sheremet2023}. However, in the weak-driving limit, where the system remains predominantly in the ground state, the quantum jump terms vanish. Consequently, the dynamics can be accurately captured by an effective non-Hermitian Hamiltonian \cite{Ashida2020} (see Supplementary). In this low-excitation regime, the Hilbert space is restricted to the many-body ground state $\ket{G} = \ket{g}^{\otimes N}$ and the single-excitation manifold $\ket{\Psi(t)}$. The collective state can then be written as
\begin{equation}
    \ket{\Psi_\text{tot}(t)}\approx \ket{G}+\ket{\Psi(t)}
    \equiv \ket{G}+\sum_{\mu=1}^{N}p_\mu(t)\ket{\psi_\mu},
\end{equation}
where $\ket{\psi_\mu}= \sigma_\mu^\dagger\ket{G}$ denotes the bare state in which only the $\mu$th emitter is excited, $\sigma_\mu^\dagger=\vert{}e_\mu\rangle\langle g_\mu\vert{}$ and $\sigma_\mu=(\sigma_\mu^\dagger)^\dagger$ are the raising and lowering operators of atom $\mu$. In the weak-excitation regime, the total single-excitation population satisfies the condition $P_\text{exc}(t)=\sum_{\mu=1}^N \vert{}p_\mu(t)\vert{}^2 \ll 1$. Under this condition, the dynamics of the state $\ket{\Psi_\text{tot}(t)}$ are governed by an effective Schrödinger equation ($\hbar=1$):
\begin{equation}
\frac{d}{dt}\ket{\Psi_\text{tot}(t)}=-iH_{\mathrm{eff}}\ket{\Psi_\text{tot}(t)},
\end{equation}
where the non-Hermitian effective Hamiltonian $H_{\mathrm{eff}} = H_\mathrm{D} + H_\mathrm{A}$ comprises an external driving term $H_\mathrm{D}$ and an effective atomic interaction term $H_\mathrm{A}$. The coherent external drive is parameterized by the site-dependent Rabi frequencies $\Omega_\mu(t)$ :
\begin{equation}
H_\mathrm{D}=\frac{1}{2}\sum_{\mu=1}^{N}\Omega_\mu(t)(\sigma_\mu+\sigma^\dagger_\mu),
\end{equation}
and the mediated atomic interaction Hamiltonian takes the form:
\begin{equation}
\begin{aligned}
H_{\mathrm{A}}(t)
=&-i\frac{\Gamma_{\mathrm{tot}}}{2}
\sum_{\mu=1}^{N}\sigma_\mu^\dagger\sigma_\mu
+\Gamma_{\mathrm{fs}}
\sum_{\mu\ne\nu}
\left(
\widetilde{J}_{\mu\nu}
-\frac{i}{2}\widetilde{\gamma}_{\mu\nu}
\right)
\sigma_\mu^\dagger\sigma_\nu\\
&-i\Gamma_{\mathrm{L}}
\sum_{\mu<\nu}
e^{ik_0 r_{\mu\nu}}\sigma_\mu^\dagger\sigma_\nu
-i\Gamma_{\mathrm{R}}
\sum_{\mu>\nu}
e^{ik_0 r_{\mu\nu}}\sigma_\mu^\dagger\sigma_\nu.
\end{aligned}
\label{eq:effective_hamiltonian}
\end{equation}
Here, $r_{\mu\nu} = \vert{}x_\mu - x_\nu\vert{}$ is the interatomic separation. The atomic positions are ordered such that $x_1 < \dots < x_N$. Both $H_\mathrm{D}$ and $H_\mathrm{A}$ are expressed in the rotating frame of the driving frequency. We consider the system in the resonance case, where the driving frequency equals the bare atomic transition frequency $\omega_0$. In this regime, the effective wave vector of the guided mode is equal to the free-space wave vector $k_0 = \omega_0/c$.

The terms in $H_{\mathrm{A}}(t)$ characterize distinct physical processes. The first diagonal term represents individual spontaneous emission, where $\Gamma_{\mathrm{tot}} = \Gamma_{\mathrm{wg}} + \Gamma_{\mathrm{fs}}$ is the total single-atom decay rate, accounting for both guided-mode ($\Gamma_{\mathrm{wg}}$) and nonguided free-space ($\Gamma_{\mathrm{fs}}$) emission. Single-atom Lamb shifts are absorbed into the bare atomic transition frequency $\omega_0$.

The second term in $H_{\mathrm{A}}$ describes the collective free-space coupling, including both coherent dipole exchange and collective decay \cite{Breuer2002}. We assume that all atomic transition dipoles are parallel to one another and perpendicular to the array axis, corresponding to $\theta=\pi/2$. This transverse-dipole configuration fixes the dipole orientation relative to the interatomic separation, and we evaluate the dimensionless free-space kernels below at this angle.
\begin{equation}
\begin{aligned}
\widetilde{J}_{\mu\nu}
=\frac{3}{4}\Bigg[
&-\frac{\cos\xi_{\mu\nu}}{\xi_{\mu\nu}}
+\frac{\sin\xi_{\mu\nu}}{\xi_{\mu\nu}^{2}}
+\frac{\cos\xi_{\mu\nu}}{\xi_{\mu\nu}^{3}}
\Bigg],
\end{aligned}
\label{eq:free_space_exchange}
\end{equation}
and
\begin{equation}
\begin{aligned}
\widetilde{\gamma}_{\mu\nu}
=\frac{3}{2}\Bigg[
&\frac{\sin\xi_{\mu\nu}}{\xi_{\mu\nu}}
+\frac{\cos\xi_{\mu\nu}}{\xi_{\mu\nu}^{2}}
-\frac{\sin\xi_{\mu\nu}}{\xi_{\mu\nu}^{3}}
\Bigg],
\end{aligned}
\label{eq:free_space_collective_decay}
\end{equation}
These kernels are evaluated for $\mu \ne \nu$, defining the dimensionless distance $\xi_{\mu\nu} = k_0 r_{\mu\nu}$. $\widetilde{J}_{\mu\nu}$ gives the coherent exchange, while $\widetilde{\gamma}_{\mu\nu}$ gives collective free-space decay.

Finally, the last two terms of $H_{\mathrm{A}}(t)$ describe waveguide-mediated interactions. The decay rates into the left- and right-propagating guided modes are $\Gamma_{\mathrm{L}}$ and $\Gamma_{\mathrm{R}}$, respectively, with $\Gamma_{\mathrm{wg}} = \Gamma_{\mathrm{L}} + \Gamma_{\mathrm{R}}$. We characterize their imbalance by the directionality parameter $\eta = (\Gamma_{\mathrm{R}} - \Gamma_{\mathrm{L}})/\Gamma_{\mathrm{wg}}$, so that $\eta=0$ describes symmetric emission and $\eta=\pm1$ describes strictly unidirectional (chiral) coupling \cite{Carmichael1993,Stannigel2012}. The waveguide coupling efficiency is $\beta = \Gamma_{\mathrm{wg}}/\Gamma_{\mathrm{tot}}$, and therefore $\Gamma_{\mathrm{fs}}=(1-\beta)\Gamma_{\mathrm{tot}}$. Equivalently, the rates appearing in $H_{\mathrm{A}}$ are $\Gamma_{\mathrm{R}}=\frac{\beta\Gamma_{\mathrm{tot}}}{2}(1+\eta)$ and $\Gamma_{\mathrm{L}}=\frac{\beta\Gamma_{\mathrm{tot}}}{2}(1-\eta)$; hence $\beta$ is included through these rates rather than appearing as a separate prefactor. Under the weak-excitation approximation, we can project the dynamics directly onto the single-excitation manifold:
\begin{equation}\ket{\dot \Psi(t)}=-iH_\mathrm{A}\ket{\Psi(t)}-\frac{i}{2}\sum_{\mu=1}^{N}\Omega_\mu(t)\ket{\psi_\mu},
\label{eq:state-time-derivative}
\end{equation}

\begin{figure*}[t]
    \centering
    \IfFileExists{Fig2.png}{%
        \includegraphics[width=1\textwidth]{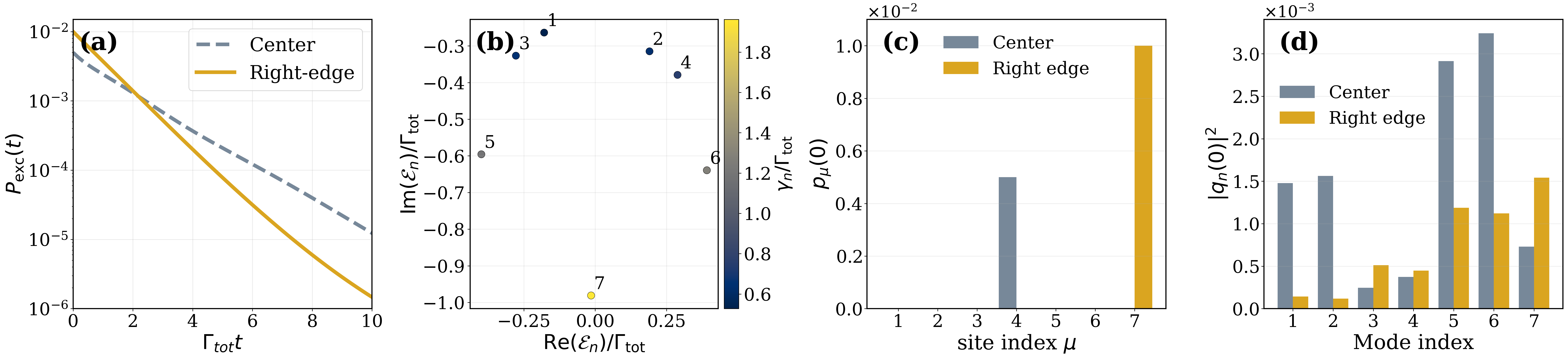}%
    }{%
        \IfFileExists{Fig1.pdf}{%
            \includegraphics[width=0.8\textwidth]{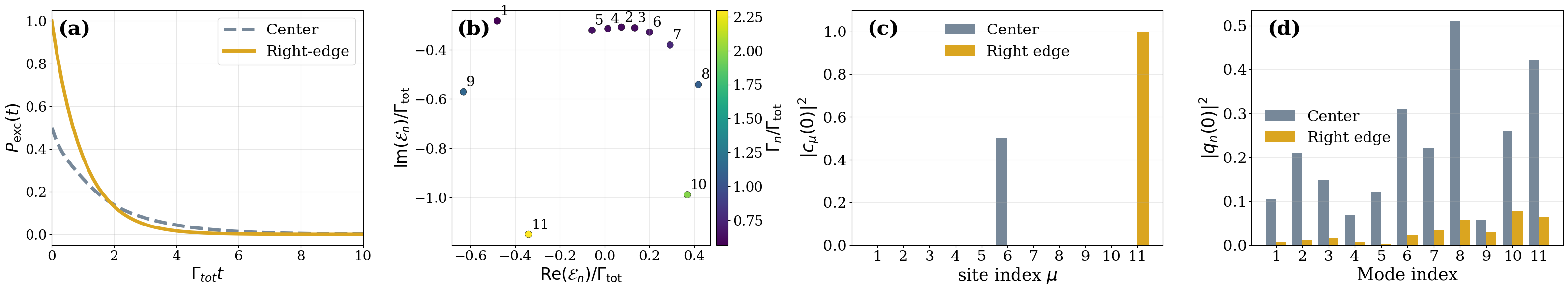}%
        }{%
            \fbox{\parbox[c][4.2cm][c]{0.92\textwidth}{\centering
            Figure 1 placeholder\\[0.5em]
            Upload the figure as \texttt{Fig1.pdf} or \texttt{Fig1.png}.}}%
        }%
    }
    \caption{Mpemba relaxation in a waveguide-coupled atomic array. Parameters are set to $N=7$, $\Gamma_{\rm tot}=1$, $\beta=0.9$, $\eta=0.75$ and $\xi=1.5\pi$. (a) Total excitation population starting from two different initial single-site excitations: the central atom excited (gray curve) and the rightmost atom excited (gold curve). (b) Complex eigenspectrum of the effective non-Hermitian Hamiltonian. Numbers label the collective eigenmodes indexed by ascending decay rates. (c) Initial-state profiles showing the site-resolved excitation population. (d) State decomposition into collective decay eigenmodes ordered by ascending decay rates.}
    \label{fig:mpemba}
\end{figure*}

To illustrate the Mpemba effect in this platform, we first consider a simple free-decay scenario. Figure~\ref{fig:mpemba}(a) demonstrates this counterintuitive relaxation by tracking the total excitation population $P_\text{exc}(t)$ over time for two distinct initial states. Although one state initially lies farther from the vacuum, it paradoxically relaxes faster. This phenomenon can be formally understood by projecting each initial state onto the eigenmodes of the effective atomic Hamiltonian $H_{\mathrm{A}}$ \cite{Ashida2020, Torres2014}. Diagonalizing the corresponding non-Hermitian matrix yields complex eigenenergies $\mathcal E_n = \omega_n - i\frac{\gamma_n}{2}$ and a biorthogonal basis of right ($\ket{\phi_n^{\mathcal R}}$) and left ($\bra{\phi_n^{\mathcal L}}$) eigenstates. Here, $\omega_n$ represents the coherent energy shift and $\gamma_n$ defines the decay rate of the population for the $n$th eigenmode. This complex eigenspectrum is displayed in Fig.~\ref{fig:mpemba}(b), illustrating the distribution of these decay rates across the collective modes.

Through this eigenmode decomposition, the driven-dissipative dynamics can be solved analytically \cite{Yu2026}. For a time-independent drive, the state evolution can be written as
\begin{equation}
\ket{\Psi(t)}=\sum_{n=1}^{N}
\left\{-\frac{\Delta_n}{2\mathcal E_n}
+e^{-i\mathcal E_n t}\left[q_n(0)+\frac{\Delta_n}{2\mathcal E_n}\right]\right\}\ket{\phi_n^{\mathcal R}},
\label{eq:constant-drive-solution}
\end{equation}
where $q_n(0)$ is the initial probability amplitude of the $n$th right eigenstate $\ket{\phi_n^{\mathcal R}}$, and $\Delta_n = \bra{\phi_n^{\mathcal L}} \sum_{\mu=1}^{N}\Omega_\mu\ket{\psi_\mu}$ quantifies the spectral overlap between the external drive and the corresponding left eigenstate. At the end of evolution, the system approaches the steady state
\begin{equation}
\ket{\Psi_{\mathrm{ss}}}
=-\sum_{n=1}^{N}\frac{\Delta_n}{2\mathcal E_n}\ket{\phi_n^{\mathcal R}}.
\end{equation}
Equation~\eqref{eq:constant-drive-solution} can also be expressed separately for each eigenmode amplitude $q_n(t)$ as
\begin{equation}
q_n(t)+\frac{\Delta_n}{2\mathcal E_n}
=e^{-i\omega_n t-\frac{\gamma_n}{2}t}
\left[q_n(0)+\frac{\Delta_n}{2\mathcal E_n}\right].
\label{eq:eigenmode-drive-solution}
\end{equation}
This form explicitly demonstrates that the deviation of each mode's instantaneous amplitude from its steady-state value, $q_n(\infty) = -\Delta_n/2\mathcal E_n$, decays exponentially in time from the initial deviation, at the decay rate of the mode.

During free decay, the drive is absent ($\Delta_n=0$), reducing the solution to $q_n(t)=e^{-i\mathcal E_n t}q_n(0)$. This result shows that a state with a greater initial weight $q_n(0)$ in the more subradiant modes (those with relatively smaller $\gamma_n$) relaxes more slowly. We can now connect this back to the free-decay example by examining the specific spatial profiles of our two initial states, one with the central atom excited and the other with the rightmost atom excited, as depicted in Fig.~\ref{fig:mpemba}(c). Figure~\ref{fig:mpemba}(d) presents the corresponding eigenmode compositions of these two initial states. Even though the state with the rightmost atom excited is initially farther from the vacuum state, its modal decomposition reveals less overlap with the slowest-decaying eigenmodes. It therefore reaches the vacuum state earlier than the state with the central atom excited, robustly demonstrating the Mpemba effect. Importantly, this phenomenon occurs widely across different parameter regimes and is not restricted to waveguide-coupled arrays. The Mpemba effect persists even when $\beta=0$, where the emitters interact purely through free-space dipole-dipole coupling (see Supplementary). The only exception arises for perfect unidirectional coupling ($\beta=1$ and $\eta=1$). Under these strictly cascaded conditions, the system reaches an exceptional point where all eigenmodes coalesce and possess identical decay rates.

\subsection*{Accelerated Reset to the Vacuum State via Active Control}
Because the Mpemba effect occurs in this system, it can be exploited to accelerate the relaxation process. Consider a situation in which, after a sequence of quantum operations, the system is left in a certain final state and must be reset to the collective ground state before the next procedure can begin. Instead of allowing the system to decay freely, we apply an external drive to the atoms during the evolution, thereby steering the original state into a configuration that decays more rapidly and resets at a faster rate.

\begin{figure*}[t]
    \centering
    \IfFileExists{Fig3_combined_2x3.pdf}{%
        \includegraphics[width=0.8\textwidth]{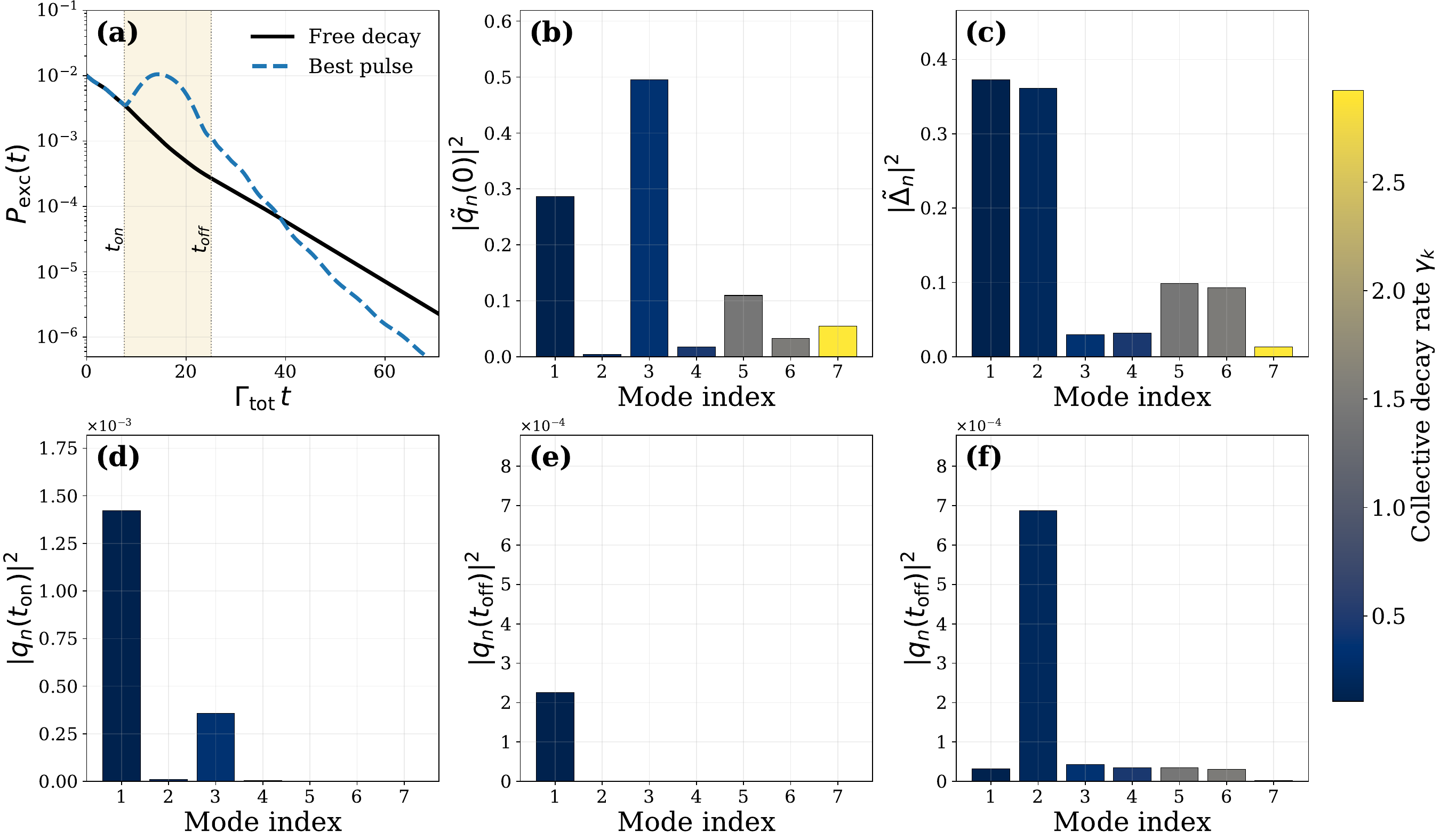}%
    }{%
        \IfFileExists{Fig2.pdf}{%
            \includegraphics[width=0.8\textwidth]{Fig1.png}%
        }{%
            \fbox{\parbox[c][6.5cm][c]{0.92\textwidth}{\centering
            Figure 1 placeholder\\[0.5em]
            Upload the figure as \texttt{Fig2.pdf} or \texttt{Fig2.png}.}}%
        }%
    }
    \caption{Accelerated Relaxation through the active control. The parameters are the same as in Fig.~\ref{fig:mpemba}. The drive is applied to atom $\mu=4$. (a) Total excitation probability \(P_{\mathrm{exc}}(t)\) under free decay (black solid line) and the optimized single-site pulse (blue dashed line). The shaded region marks the drive interval, with \(\Gamma_{\mathrm{tot}}t_{\mathrm{on}}\approx7.60\) and \(\Gamma_{\mathrm{tot}}t_{\mathrm{off}}\approx 25.02\). (b) Initial modal coefficient weights, \(|\tilde{q}_n(0)|^2=|q_n(0)|^2/\sum_m|q_m(0)|^2\). \textcolor{black}{(c) Normalized drive overlaps, $\tilde{\Delta}_n=\Delta_n/{\sqrt{\sum_m|\Delta_m|^2}}$.} (d) Modal weights at pulse onset, \(t_{\mathrm{on}}\), when the driven and freely evolving states coincide. (e,f) Modal weights at \(t_{\mathrm{off}}\) for free decay and the driven protocol, respectively, illustrating the drive-induced suppression of the most subradiant contribution. Modes are ordered by increasing collective decay rate \(\gamma_n\), with \(n=1\) denoting the most subradiant mode; bar colors indicate \(\gamma_n\). The snapshot weights in (d–f) are unnormalized, and (e,f) share the same vertical scale.}
    \label{fig:relaxation}
\end{figure*}

In Fig.~\ref{fig:relaxation}(a), the system is initially prepared in the Dicke state \cite{Dicke1954}. Compared with free decay, applying a constant drive to a single site for a finite period (turned on at $t_{\mathrm{on}}$ and off at $t_{\mathrm{off}}$) effectively accelerates the relaxation process. This acceleration can be understood through spectral analysis. The initial Dicke state has a non-negligible overlap with the most subradiant eigenstate, which is the mode with the smallest decay rate, as shown in Fig.~\ref{fig:relaxation}(b). Due to this small decay rate, the corresponding component evolves on a long time scale during free decay. However, when the drive is applied at an appropriate time, its component in the most subradiant mode [Fig.~\ref{fig:relaxation}(c)] interferes destructively with the corresponding component of the system. This operation effectively redistributes the spectral weight among modes with higher decay rates. Figure~\ref{fig:relaxation}(d) displays the eigenspace profile of the state at the turn-on time $t_{\mathrm{on}}$, while Fig.~\ref{fig:relaxation}(e,f) compares the profiles of the freely decaying and actively driven states at the turn-off time $t_{\mathrm{off}}$. As observed, the weight of the most subradiant mode is dramatically reduced by the transient drive compared with the free-decay baseline. The accompanying increase in trace distance results from the enhanced weights of the other eigenmodes; however, because these alternative modes possess larger decay rates, they dissipate on a shorter time scale than the most subradiant mode, leading to an accelerated relaxation to the vacuum state.


This active-control protocol therefore provides an effective strategy for accelerating the state reset process. Its effectiveness is not limited to driving atom 4 as shown in Fig.~\ref{fig:relaxation}: driving any of the other atoms also accelerates relaxation under the same parameters, as demonstrated in Table~\ref{tab:all_sites_relaxation}. Consistent with our observations for free decay, this active-control strategy is highly robust across different parameter regimes (see Supplementary).  Specifically, the acceleration protocol remains effective even in the absence of a waveguide ($\beta=0$).

\begin{table}[t]
\centering
\begin{tabular*}{\columnwidth}{@{\extracolsep{\fill}}lrrrr@{}}
\toprule
Driven atom
& $t_{\mathrm{on}}$
& $t_{\mathrm{off}}$
& $t_{\mathrm{th}}$
& $\Delta t$ (\%) \\
\midrule
Free decay & ---      & ---      & 56.706 &  0.00 \\
1          & 14.369 & 30.111 & 51.985 &  8.33 \\
2          & 11.601 & 28.536 & 50.936 & 10.18 \\
3          &  9.499 & 26.824 & 51.551 &  9.09 \\
\textbf{4}          &  \textbf{7.597} & \textbf{25.016} & \textbf{48.444} & \textbf{14.57} \\
5          &  5.908 & 23.275 & 43.055 & 24.07 \\
6          &  4.586 & 21.499 & 42.836 & 24.46 \\
7          &  4.016 & 36.639 & 46.250 & 18.44 \\
\bottomrule
\end{tabular*}
\caption{
Site dependence of the single-site driving protocol.
For each driven atom, $t_{\mathrm{on}}$ is the first instant
at which the local contribution of the most subradiant mode
is opposite in phase to the drive, and $t_{\mathrm{off}}$
is optimized to minimize the first-passage time
$t_{\mathrm{th}}$ to $P_{\mathrm{exc}}\leq10^{-5}$.
The drive amplitude and phase are identical for all sites.
The relative time reduction is defined as
$\Delta t(\%)=100(t_{\mathrm{ref}}-t_{\mathrm{th}})/t_{\mathrm{ref}}$,
where $t_{\mathrm{ref}}$ is the free-decay threshold time.
For the parameters considered, driving any of the seven atoms
accelerates relaxation, with atom 6 yielding the largest
time reduction.
Times are expressed in units of $\Gamma_{\mathrm{tot}}^{-1}$. The protocol illustrated in Fig.~\ref{fig:relaxation}(a)--(f)
corresponds to driving atom 4.
}
\label{tab:all_sites_relaxation}
\end{table}



\subsection*{Accelerating State Preparation through Two-Step Driving Protocol}
We now turn to the task of state preparation. Starting from Eq.~\eqref{eq:eigenmode-drive-solution}, we consider the case in which the system is initially in the vacuum state, for which the modal amplitudes satisfy
\begin{equation}
q_n(t)+\frac{\Delta_n}{2\mathcal E_n}
=e^{-i\mathcal E_n t}
\frac{\Delta_n}{2\mathcal E_n}.
\label{eq:driven-from-vacuum}
\end{equation}
We choose the target state to be
\begin{equation}
\ket{\Psi_{\mathrm{target}}}
=-\sum_{n=1}^{N}\frac{\Delta_n'}{2\mathcal E_n}\ket{\phi_n^{\mathcal R}},
\end{equation}
where $\Delta_n'$ denotes the modal component of the drive for which the target state is the steady state. The difference between the target and current modal amplitudes is then
\begin{equation}
q_n(t)+\frac{\Delta_n'}{2\mathcal E_n}
=\frac{1}{2\mathcal E_n}
\left[\Delta_n'+\left(e^{-i\mathcal E_n t}-1\right)\Delta_n\right].
\label{eq:target-state-deviation}
\end{equation}
This expression provides a crucial insight: if the initial transient drive possesses a greater weight in the more subradiant modes, it can propel these slowly evolving components toward their target values more rapidly, thereby accelerating the overall preparation process. After an appropriate switching time, however, the drive must be changed back to the one for which the desired target state is the steady state.

\begin{figure*}[t]
    \centering
    \IfFileExists{Fig4.pdf}{%
        \includegraphics[width=0.8\textwidth]{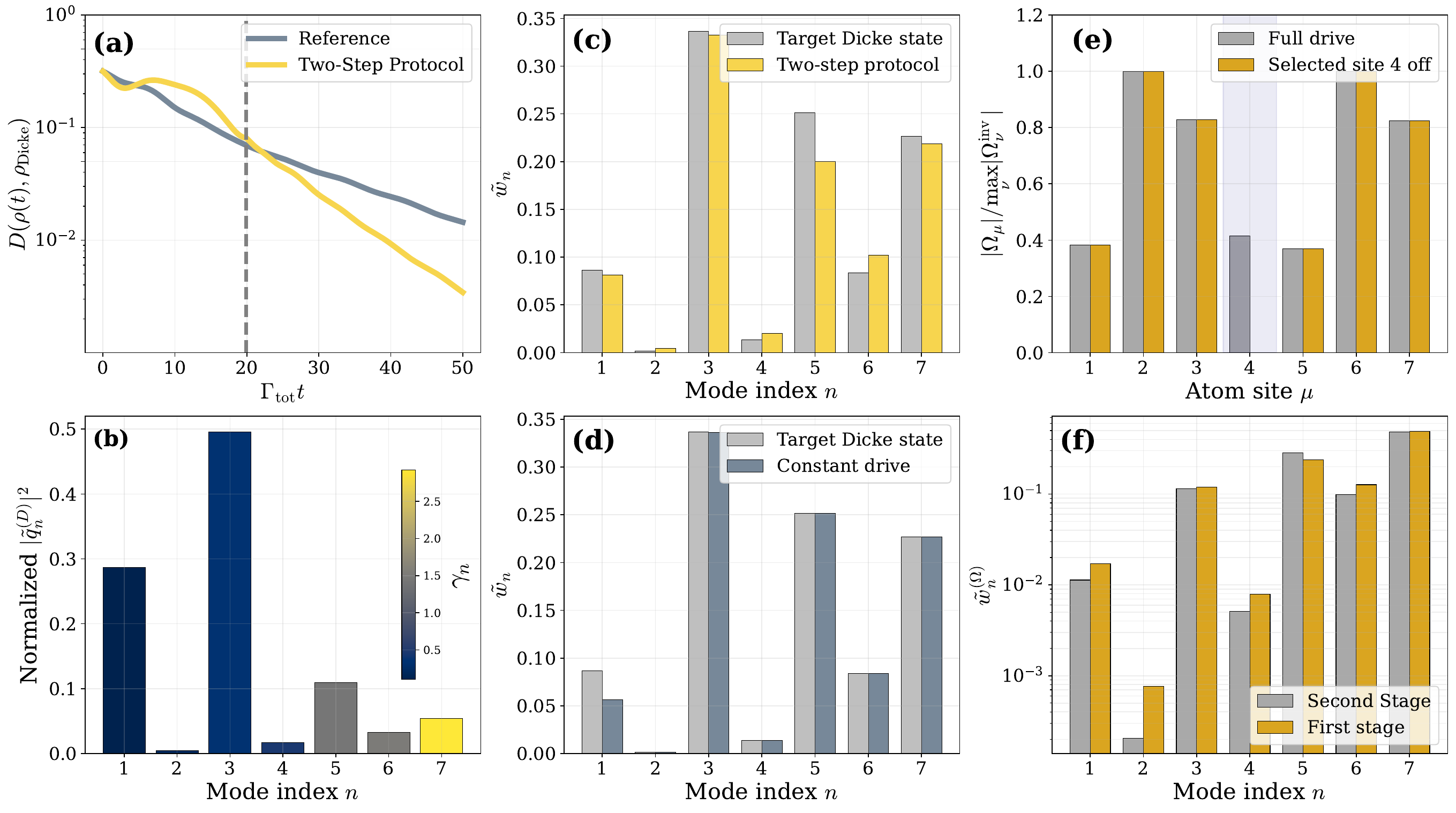}%
    }{%
        \IfFileExists{Fig3.pdf}{%
            \includegraphics[width=\textwidth]{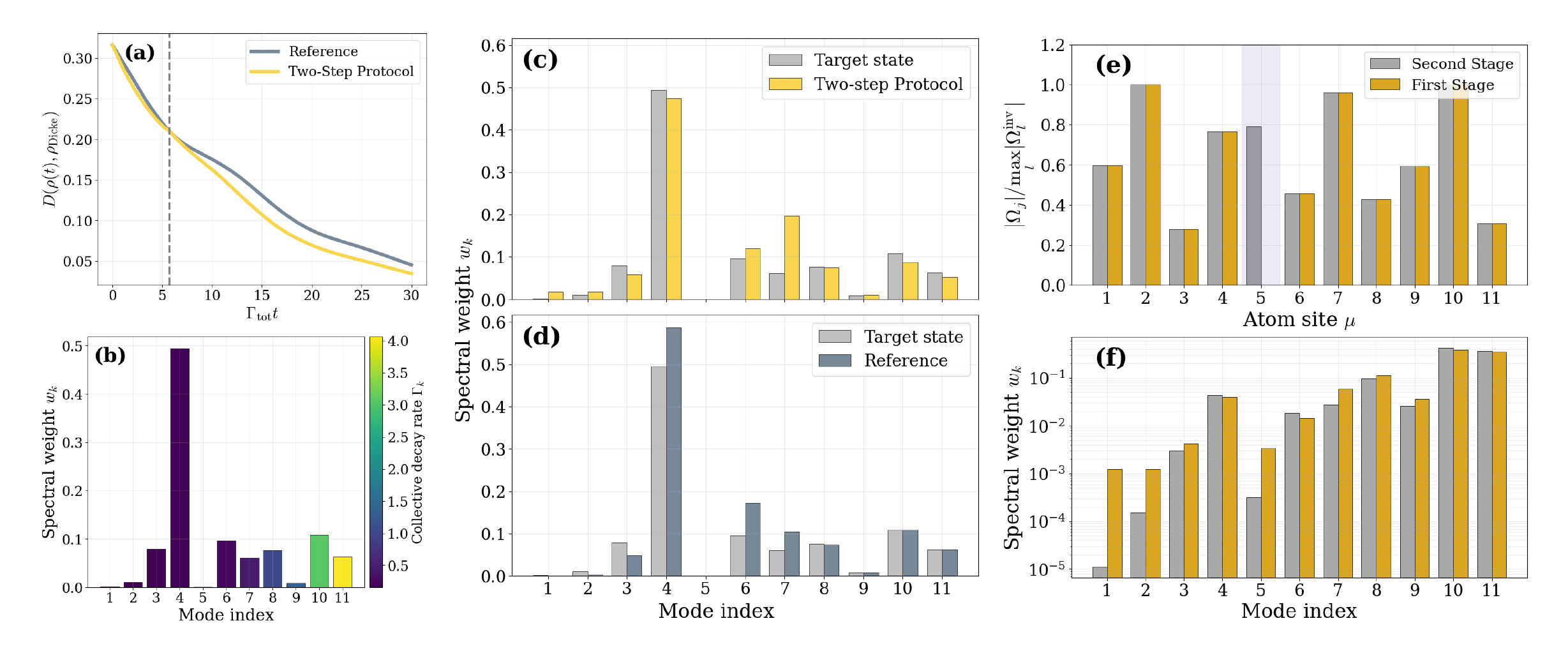}%
        }{%
            \fbox{\parbox[c][6.5cm][c]{0.92\textwidth}{\centering
            Figure 1 placeholder\\[0.5em]
            Upload the figure as \texttt{Fig3.pdf} or \texttt{Fig3.png}.}}%
        }%
    }
    \caption{Accelerated Dicke-state preparation using an energy-efficient two-step driving protocol. The parameters are the same as in Fig.~\ref{fig:mpemba}. Atom $\mu=4$ is left undriven during the first stage and switched on at the optimized switching time $t'\approx19.89$. \textcolor{black}{The trace distance is defined as $D(\rho,\sigma)=\frac{1}{2}\|\rho-\sigma\|_1$, where $\|\cdot\|_1$ denotes the trace norm.}
    (a) Time evolution of the trace distance to the target Dicke state, comparing the two-step protocol (yellow line) with the conventional constant-drive approach (gray line). The vertical dashed line marks the switching time $t'$. 
    \textcolor{black}{(b) Normalized target Dicke state modal weights $|\tilde q_n^{(D)}|^2$, where $q_n^{(D)}=\langle\phi_n^L|D\rangle$ and $\tilde q_n^{(D)}=q_n^{(D)}/\sqrt{\sum_m|q_m^{(D)}|^2}$. Modes are ordered by ascending decay rates $\gamma_n$.}
    \textcolor{black}{(c, d) State decompositions at time $t'$ for (c) the two-step state $\ket{\Psi_{\rm protocol}(t')}$ and (d) the constant-drive state $\ket{\Psi_{\rm CW}(t')}$, shown using the target Dicke state modal component weights $\tilde{w}_n(t')=\|q_n(t')\ket{\phi_n^R}\|^2/
{\sum_m\|q_m^{({D})}\ket{\phi_m^R}\|^2},
\
q_n(t')=\langle\phi_n^L|\Psi_{\rm protocol/CW}(t')\rangle$ (e) Spatial drive configurations of the two-step protocol, with the selected site $\mu=4$ undriven during the first stage. (f) Normalized spectral drive weights $\tilde{w}_n^{(\Omega)}=\|\ket{\phi_n^R}\langle\phi_n^L|\Omega\rangle\|^2/{\sum_m\|\ket{\phi_m^R}\langle\phi_m^L|\Omega\rangle\|^2}$ for the second-stage and first-stage drives.}
    }
    \label{fig:state-preparation}
\end{figure*}

\textcolor{black}{It is worth noting that} acceleration does not necessarily require supplying more energy through stronger driving: in an open quantum system, leaving selected sites undriven can reduce the total driving energy while still expediting state preparation. Figure~\ref{fig:state-preparation}(a) illustrates this by tracking the relaxation dynamics toward the target Dicke state, comparing our two-step driving protocol against a continuous constant-drive approach. This target Dicke state possesses a specific eigenmode decomposition, which is detailed in Fig.~\ref{fig:state-preparation}(b). By examining the internal state of the system at the optimal switching time $t'$, we observe distinct modal profiles for the two-step state $\ket{\Psi_1(t')}$ [Fig.~\ref{fig:state-preparation}(c)] and the constant-drive state $\ket{\Psi_2(t')}$ [Fig.~\ref{fig:state-preparation}(d)]. The underlying mechanism driving this difference is revealed by examining the spatial drive configurations in Fig.~\ref{fig:state-preparation}(e): while the constant drive uniformly stabilizes the target Dicke state (gray bars), the first stage of the two-step protocol deliberately leaves one site undriven (gold bars). As verified by the spectral drive decomposition in Fig.~\ref{fig:state-preparation}(f), this local deactivation effectively redistributes and concentrates spectral weight into the more subradiant modes. Consequently, the subradiant components of the two-step state $\ket{\Psi_1(t')}$ are driven much closer to their target values than those of the constant-drive state $\ket{\Psi_2(t')}$. Although atom 4 is utilized as the representative example in Fig.~\ref{fig:state-preparation}, this protocol successfully accelerates state preparation even when different atoms are selected to remain undriven, as documented in Table~\ref{tab:single_atom_threshold_arrival}. This counterintuitive two-step protocol therefore secures the dual advantages of reduced energy expenditure and faster state preparation. Results for other parameter sets, including the no-waveguide limit, are provided in the Supplementary Information.

\begin{table}[t]
    \centering
    \begin{tabular*}{\columnwidth}{@{\extracolsep{\fill}}lrrr}
        \toprule
        Controlled Atom & $t'$ & $t_{\rm th}$ & $\Delta t$ (\%) \\
        \midrule
        Reference & --      & 41.524 & 0.00 \\
        1                   & 15.417  & 32.989 & -20.56 \\
        2                   & 14.074  & 37.064 & -10.74 \\
        3                   &  5.715  & 38.300 &  -7.76 \\
        \textbf{4}          & \textbf{19.887} & \textbf{31.283} & \textbf{-24.66} \\
        5                   &  9.574  & 22.841 & -44.99 \\
        6                   &  3.455  & 26.338 & -36.57 \\
        7                   & 17.606  & 36.708 & -11.60 \\
        \bottomrule
    \end{tabular*}
    \caption{Optimized turn-on times and threshold-arrival times for the
    constant-drive reference and independently optimized single-atom protocols under the same parameter setting in Fig~\ref{fig:state-preparation}. For each atom, $t'$ is optimized to minimize the first
    arrival time at the normalized target-error threshold of $0.005$.
    The percentage difference is calculated as
    $\Delta t (\%)=100(t_{\rm atom}-t_{\rm ref})/t_{\rm ref}$; negative values indicate
    faster arrival than the constant-drive reference.}
    \label{tab:single_atom_threshold_arrival}
\end{table}



\subsection*{Chirality-Guided Site Selection and the non-Hermitian Skin-Mpemba Effect}

\begin{figure*}[t]
    \centering
    \IfFileExists{Fig5.pdf}{%
        \includegraphics[width=\textwidth]{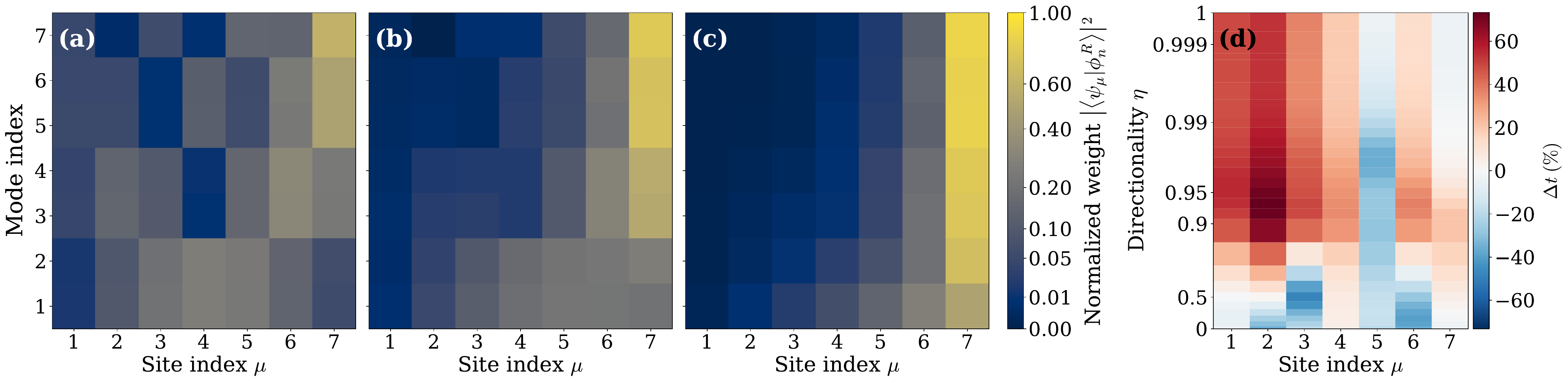}%
    }{%
        \IfFileExists{Fig2.pdf}{%
            \includegraphics[width=0.8\textwidth]{Fig1.png}%
        }{%
            \fbox{\parbox[c][6.5cm][c]{0.92\textwidth}{\centering
            Figure 1 placeholder\\[0.5em]
            Upload the figure as \texttt{Fig2.pdf} or \texttt{Fig2.png}.}}%
        }%
    }
    \caption{Collective eigenmode distributions and site-dependent preparation-time changes. (a–c) Spatial weights of the right eigenmodes for \(\eta=0.5,\ 0.9,\) and \(0.99\), respectively, at directionality. Each eigenmode is normalized such that its weights sum to unity over all sites. Modes are ordered by increasing decay rate. (d) Relative preparation-time change, \(100(t_{\mathrm{protocol}}-t_{\mathrm{CW}})/t_{\mathrm{CW}}\), versus directionality and the site whose drive is initially switched off and restored at \(t_{\mathrm{on}}=7\), for \(\beta=0.9\). The constant-drive reference is evaluated separately at each directionality, using the same normalized target-error threshold of \(0.005\). Negative values indicate faster preparation. The directionality axis is stretched near \(\eta=1\). The remaining parameters are the same as in Fig.~\ref{fig:mpemba}.}
    \label{fig:heatmaps}
\end{figure*}

The processes described above, for both state preparation and relaxation, are not restricted to waveguide-coupled systems or even to chiral environments. However, in this section, we highlight the specific benefits of introducing chirality into the system. In non-Hermitian systems, the skin effect dictates that if the system exhibits chirality (a preferred direction of interactio), the right eigenmodes will tend to localize at the preferred boundary \cite{Lee2016,Song2019,Wang2022}. In our system, this directional preference is governed by the chiral waveguide. Figure~\ref{fig:heatmaps}(a) shows the right eigenmode profiles of the effective atomic Hamiltonian $H_\text{A}$ for a moderate directionality of $\eta=0.5$. As this directionality increases to $\eta=0.9$ [Fig.~\ref{fig:heatmaps}(b)] and approaches strict chirality at $\eta=0.99$ [Fig.~\ref{fig:heatmaps}(c)], these modes become increasingly localized at the right boundary, distinctly demonstrating the non-Hermitian skin effect. The presence of the non-Hermitian skin effect in our system provides guidance on which localized initial states are likely to exhibit the Mpemba effect, as well as which local operations induce the Pontus-Mpemba effect. We term this fundamental correlation between the Mpemba relaxation dynamics and the non-Hermitian skin effect the ``non-Hermitian Skin-Mpemba effect.'' 

Applying this principle to the two-step driving protocol, we demonstrate that local operations executed on the chirally preferred side are far more likely to yield a preparation-time advantage. As shown in Fig.~\ref{fig:heatmaps}(d), as $\eta$ increases, it becomes progressively more difficult for local operations on the left side of the array to achieve any time reduction. For high directionality ($\eta>0.9$) under this parameter set, the operations that successfully accelerate Dicke-state preparation are exclusively located on the right side of the array. It should be noted that the exact site required to achieve a time advantage is not universally fixed; it depends on the chosen parameter set (such as $\xi$ and $N$) as well as the target state. However, the overarching physical trend remains robust: the Mpemba effect predominantly emerges on the side favored by the system's chirality. Results for other parameter sets and different target states are provided in the Supplementary. Ultimately, when designing protocols that utilize the Mpemba effect to accelerate state preparation, one must account for the system's chirality due to this non Hermitian Skin-Mpemba effect. By tuning the chirality, we can strategically select the advantageous operation sites.


\section*{Discussion}
This work demonstrates the distinct advantage of utilizing the Mpemba effect to accelerate both quantum state preparation and state reset processes, and further establishes that the preferred selection of local operation sites is governed by the non-Hermitian Skin-Mpemba effect. Although we consider a specific waveguide-coupled atomic array, this control philosophy should be applicable to other open quantum systems. Because the eigenmodes of such systems inherently possess varying decay rates, any operation that redistributes the spectral profile of the state will induce the Pontus-Mpemba effect. For the same reason, the non-Hermitian Skin-Mpemba effect should also be universal across diverse platforms wherever chirality exists. We hope this work encourages the community to investigate the correlation between the non-Hermitian skin effect and the Mpemba effect, not merely to observe these physical phenomena, but because this correlation is a highly valuable resource for designing practical applications. 


\section*{Method}

\subsection*{Determination of the Pulse Timing for the Resetting Protocol}

We consider a rectangular pulse applied to atom $\mu$,
with fixed amplitude $\Omega_{\mathrm d}$ and phase
$\phi_{\mathrm d}$ ($\phi_{\mathrm d}=0$ in
Fig.~\ref{fig:relaxation}). The drive on each atom
$\nu=1,\ldots,N$ is
\begin{equation}
\Omega_\nu(t)
=\delta_{\nu\mu}\Omega_{\mathrm d}e^{i\phi_{\mathrm d}}
\left[
\Theta(t-t_{\mathrm{on}})
-\Theta(t-t_{\mathrm{off}})
\right].
\end{equation}
The pulse timing is determined in two steps:
a phase-based selection of $t_{\mathrm{on}}$, followed
by numerical optimization of $t_{\mathrm{off}}$.

Let $H_{\mathrm A}$ denote the undriven effective
Hamiltonian in the single-excitation subspace.
Its right and left eigenvectors satisfy
\begin{equation}
\begin{aligned}
H_{\mathrm A}\ket{\phi_n^{\mathcal R}}
&=\mathcal E_n\ket{\phi_n^{\mathcal R}},\\
\bra{\phi_n^{\mathcal L}}H_{\mathrm A}
&=\mathcal E_n\bra{\phi_n^{\mathcal L}},\\
\braket{\phi_n^{\mathcal L}|\phi_m^{\mathcal R}}
&=\delta_{nm}.
\end{aligned}
\end{equation}
Writing $\mathcal E_n=\omega_n-i\gamma_n/2$, we order
the modes by increasing $\gamma_n$ and select the most
subradiant mode, $n=1$, independently of its initial
coefficient weight. The initial modal coefficients are
$q_n(0)=\braket{\phi_n^{\mathcal L}|\Psi(0)}$,
where the left eigenvectors have support only in the
single-excitation subspace.

Before the pulse, the local contribution of mode 1
at atom $\mu$ is
\begin{equation}
a_{\mu,1}(t)
=q_1(0)\braket{\psi_\mu|\phi_1^{\mathcal R}}
e^{-i\mathcal E_1t}.
\end{equation}
We choose the earliest nonnegative time at which this
contribution is opposite in phase to the applied drive:
\begin{equation}
t_{\mathrm{on}}^{(\mu)}
=\min\{t\geq0:
\operatorname{mod}(
\arg(a_{\mu,1}(t))-\phi_{\mathrm d},2\pi)=\pi\}.
\label{eq:turn_on_condition}
\end{equation}
Here, $\operatorname{mod}(x,2\pi)$ returns a value
in $[0,2\pi)$. Defining the initial local phase as
\begin{equation}
\vartheta_\mu
=\arg\!\left(
q_1(0)\braket{\psi_\mu|\phi_1^{\mathcal R}}
\right),
\end{equation}
the turn-on time for $\omega_1\ne0$ is
\begin{equation}
t_{\mathrm{on}}^{(\mu)}
=
\frac{
\operatorname{mod}\!\left(
\operatorname{sgn}(\omega_1)
(\vartheta_\mu-\phi_{\mathrm d}-\pi),2\pi
\right)
}{|\omega_1|}.
\label{eq:turn_on_time}
\end{equation}
This prescription requires a nonzero local modal
contribution. If $\omega_1=0$, the phase condition is
either satisfied initially or is never reached.

For each candidate switch-off time
$\tau\geq t_{\mathrm{on}}^{(\mu)}$, we construct a
trajectory consisting of free evolution before
$t_{\mathrm{on}}^{(\mu)}$, driven evolution up to
$\tau$, and free evolution thereafter.
For brevity, we write
$t_{\mathrm{on}}=t_{\mathrm{on}}^{(\mu)}$ below.
Specializing the modal solution in Eq.~(8) to the
constant single-site drive gives, for
$t_{\mathrm{on}}\leq t\leq\tau$,
\begin{equation}
\begin{aligned}
q_n(t)
={}&e^{-i\mathcal E_n(t-t_{\mathrm{on}})}
q_n(t_{\mathrm{on}})
\\
&-\frac{\Omega_{\mathrm d}e^{i\phi_{\mathrm d}}
\braket{\phi_n^{\mathcal L}|\psi_\mu}}
{2\mathcal E_n}
\left[
1-e^{-i\mathcal E_n(t-t_{\mathrm{on}})}
\right],
\end{aligned}
\label{eq:driven_modal_evolution}
\end{equation}
where
$q_n(t_{\mathrm{on}})
=q_n(0)e^{-i\mathcal E_nt_{\mathrm{on}}}$,
and the continuous limit is used when
$\mathcal E_n=0$.
After switch-off, the coefficients evolve freely as
\begin{equation}
q_n(t)=q_n(\tau)e^{-i\mathcal E_n(t-\tau)},
\qquad t\geq\tau.
\end{equation}
The total excitation probability for this trajectory is
\begin{equation}
P_{\mathrm{exc}}(t;\tau)
=
\left\|
\sum_n q_n(t;\tau)\ket{\phi_n^{\mathcal R}}
\right\|^2.
\end{equation}

For the results presented here, the initial state is
\begin{equation}
\ket{\Psi(0)}
=\sqrt{1-\epsilon}\ket{G}
+\sqrt{\epsilon}e^{i\phi_0}\ket{D},
\qquad
\epsilon=0.01,\quad \phi_0=0,
\end{equation}
where $\ket{D}$ is the normalized single-excitation
Dicke state. Thus, $P_{\mathrm{exc}}(0)=\epsilon=0.01$.
We use the absolute threshold $P_{\mathrm{th}}=10^{-5}$,
corresponding to $10^{-3}$ of the initial excitation
probability, and define the first-passage time as
\begin{equation}
T_\mu(\tau)
=\inf\{t\geq0:
P_{\mathrm{exc}}(t;\tau)\leq P_{\mathrm{th}}\}.
\end{equation}
The switch-off time is selected according to
\begin{equation}
t_{\mathrm{off}}^{(\mu)}
=\underset{
\tau\in[t_{\mathrm{on}}^{(\mu)},\,t_{\mathrm{ref}}]
}{\operatorname{arg\,min}}
T_\mu(\tau),
\label{eq:optimal_switch_off}
\end{equation}
where $t_{\mathrm{ref}}$ is the threshold time under
free decay from the same initial state.
The zero-duration pulse,
$\tau=t_{\mathrm{on}}^{(\mu)}$, is included as the
free-decay baseline.
If the reference reaches the threshold before turn-on,
no acceleration is possible with this prescribed
turn-on time.

Numerically, we first scan 257 uniformly spaced
switch-off times and refine each sampled local minimum
using bounded scalar minimization.
Threshold crossings are located using Brent's
root-finding method.
If continuous driving reaches the threshold within the
search interval, its first crossing is included as a
candidate switch-off time and bounds the remaining
search.
The scan density is successively increased as
$M\rightarrow2M-1$ until the optimal first-passage time
changes by less than $10^{-4}\Gamma_{\mathrm{tot}}^{-1}$
for two consecutive refinements, with a maximum of
seven scan levels.

\subsection*{Analytical Conditions and Numerical Search for the Optimal Turn-On Time}
\label{subsec:turn-on-time}

We consider a protocol in which a selected atom $\mu\in\{1,\ldots,N\}$
is undriven during the first stage and is switched on at time $t'$. The
drive vector is
\begin{equation}
    \ket{\Omega(t)}
    =
    \sum_{\nu=1}^{N}\Omega_\nu(t)\ket{\psi_\nu},
\end{equation}
where $\ket{\psi_\nu}$ is the single-excitation state with atom $\nu$
excited. During the first stage, $\Omega_\mu(t)=0$, while all other
atoms are driven by their inverse-designed amplitudes. At $t=t'$, the
drive on atom $\mu$ is restored, so that all atoms are driven by the
inverse-designed field.

Following Eq.~\ref{eq:state-time-derivative}, the state in the single-excitation sector obeys
\begin{equation}
\begin{aligned}
    \frac{d}{dt}\ket{\Psi(t)}
    &=
    -iH_{\mathrm A}\ket{\Psi(t)}
    -\frac{i}{2}\ket{\Omega(t)}
\end{aligned}
\label{eq:single-excitation-dynamics}
\end{equation}
where $H_{\mathrm A}$ is the effective non-Hermitian Hamiltonian.

The left and right eigenvectors of $H_{\mathrm A}$ and their
biorthogonal normalization are defined in Eq.~(15). We denote their
eigenvalues by $\mathcal E_n$. The inverse-designed drive makes
$\ket{\Psi_{\mathrm{target}}}$ a steady state of Eq.~\eqref{eq:single-excitation-dynamics};
therefore,
\begin{equation}
    \ket{\Omega^{\mathrm{inv}}}
    =
    -2H_{\mathrm A}\ket{\Psi_{\mathrm{target}}}.
\end{equation}
Decompose this drive into contributions from individual sites:
\begin{equation}
\begin{aligned}
    \ket{\Omega^{\mathrm{inv}}}
    &= \sum_{\nu=1}^{N}\ket{\Omega_\nu},\\
    \ket{\Omega_\nu}
    &= (\Omega^{\mathrm{inv}})_\nu\ket{\psi_\nu}.
\end{aligned}
\end{equation}
Here, $(\Omega^{\mathrm{inv}})_\nu$ is the inverse-designed drive
amplitude at site $\nu$. The first-stage drive omits the selected
site's contribution:
\begin{equation}
    \ket{\Omega^{(1)}_\mu}
    =
    \sum_{\substack{\nu=1\\\nu\ne\mu}}^{N}\ket{\Omega_\nu}
    =
    \ket{\Omega^{\mathrm{inv}}}-\ket{\Omega_\mu}.
\end{equation}

Define the modal drive
amplitudes
\begin{equation}
\begin{aligned}
    \Delta'_{n,\nu}
    &= \langle\phi_n^{\mathcal L}|\Omega_\nu\rangle,\\
    \Delta'_n
    &= \langle\phi_n^{\mathcal L}|\Omega^{\mathrm{inv}}\rangle
     = \sum_{\nu=1}^{N}\Delta'_{n,\nu}.
\end{aligned}
\end{equation}
The response of mode $n$ to a constant drive over a duration $u$ is
\begin{equation}
    f_n(u)
    =
    \frac{e^{-i\mathcal E_n u}-1}{\mathcal E_n}.
\end{equation}

For a selected site $\mu$ and switching time $t'$, the biorthogonal
modal amplitude $q_n(t;t',\mu)$ is
\begin{equation}
\begin{cases}
    q_n(t;t',\mu)
    =
    \dfrac{1}{2}
    \bigl(\Delta'_n-\Delta'_{n,\mu}\bigr)f_n(t),
    & 0\leq t\leq t',\\[6pt]
    q_n(t;t',\mu)
    =
    \dfrac{1}{2}
    \bigl(\Delta'_n-\Delta'_{n,\mu}\bigr)f_n(t')
    e^{-i\mathcal E_n(t-t')}
    \\[2pt]
    \qquad{}+
    \dfrac{1}{2}\Delta'_n f_n(t-t'),
    & t>t'.
\end{cases}
\label{eq:delayed-site-modal-state}
\end{equation}
Define the modal-amplitude vector
\begin{equation}
    \mathbf q_\mu(t,t')
    =
    \bigl(q_1(t;t',\mu),\ldots,q_N(t;t',\mu)\bigr)^{\mathsf T}.
\end{equation}
The corresponding excitation-sector state is
\begin{equation}
    \ket{\Psi_\mu(t;t')}
    =
    \sum_{n=1}^{N}q_n(t;t',\mu)\ket{\phi_n^{\mathcal R}}.
\end{equation}

We measure the preparation error by the normalized squared distance
in the single-excitation sector:
\begin{equation}
\begin{aligned}
    \varepsilon_\mu(t,t')
    &=
    \frac{
      \left\|
        \ket{\Psi_\mu(t;t')}
        -\ket{\Psi_{\mathrm{target}}}
      \right\|^2
    }{
      P_{\mathrm{target}}
    },\\
    \ket{\Psi_{\mathrm{target}}}
    &= \sqrt{P_{\mathrm{target}}}\ket{\mathrm{Dicke}},
\end{aligned}
\label{eq:normalized-target-error}
\end{equation}
where $\ket{\mathrm{Dicke}}$ is the normalized target Dicke state and
$P_{\mathrm{target}}=
\langle\Psi_{\mathrm{target}}|\Psi_{\mathrm{target}}\rangle$ is its
excitation population ($P_{\mathrm{target}}=0.1$ for Fig.~\ref{fig:state-preparation}). In the modal basis, define
\begin{equation}
\begin{aligned}
    \mathbf q_{\mathrm{target}}
    &=
    L^\dagger\ket{\Psi_{\mathrm{target}}},\\
    G_{nm}
    &=
    \langle\phi_n^{\mathcal R}|\phi_m^{\mathcal R}\rangle.
\end{aligned}
\end{equation}
Then the error can equivalently be written as
\begin{equation}
    \varepsilon_\mu(t,t')
    =
    \frac{
      \bigl(\mathbf q_\mu(t,t')-\mathbf q_{\mathrm{target}}\bigr)^\dagger
      G
      \bigl(\mathbf q_\mu(t,t')-\mathbf q_{\mathrm{target}}\bigr)
    }{
      P_{\mathrm{target}}
    }.
\end{equation}
The Gram matrix $G$ accounts for the non-orthogonality of the right
eigenvectors.

For an interior switching-time solution with $t>t'$, differentiating
Eq.~\eqref{eq:delayed-site-modal-state} gives
\begin{equation}
    \frac{\partial\ket{\Psi_\mu(t;t')}}{\partial t'}
    =
    \frac{i}{2}
    \sum_{n=1}^{N}
       \Delta'_{n,\mu}e^{-i\mathcal E_n(t-t')}
       \ket{\phi_n^{\mathcal R}}.
\end{equation}
Consequently,
\begin{equation}
    \frac{\partial\varepsilon_\mu(t,t')}{\partial t'}
    =
    \frac{2}{P_{\mathrm{target}}}
    \operatorname{Re}\!\left[
      \bigl(\ket{\Psi_\mu(t;t')}-\ket{\Psi_{\mathrm{target}}}\bigr)^\dagger
      \frac{\partial\ket{\Psi_\mu(t;t')}}{\partial t'}
    \right].
\label{eq:error-delay-derivative}
\end{equation}

For Fig.~\ref{fig:state-preparation} and Table~\ref{tab:single_atom_threshold_arrival}, the target-error threshold is
$\varepsilon_{\mathrm{th}}=0.005$. For each site $\mu$, let $t_{\max}$
denote the maximum simulation time. The earliest attainable threshold
time is
\begin{equation}
    T_\mu
    =
    \inf\left\{
      T\in[0,t_{\max}]:
      \min_{0\leq t'\leq T}\varepsilon_\mu(T,t')
      \leq\varepsilon_{\mathrm{th}}
    \right\}.
\label{eq:site-threshold-time}
\end{equation}
If the threshold is not reached for any $T\leq t_{\max}$, we set
$T_\mu=\infty$ and report that the threshold was not reached within the
simulation window.

At an interior optimum $(T_\mu,t_\mu^{\prime\star})$, the threshold and
stationarity conditions are
\begin{align}
    \varepsilon_\mu(T_\mu,t_\mu^{\prime\star})
    &= \varepsilon_{\mathrm{th}},
    \label{eq:threshold-condition}\\
    \left.
    \frac{\partial\varepsilon_\mu(t,t')}{\partial t'}
    \right|_{(T_\mu,t_\mu^{\prime\star})}
    &=0.
    \label{eq:stationarity-condition}
\end{align}
We also evaluate the boundary cases $t'=0$, corresponding to
continuous driving, and $t'=T$, where the selected site remains
undriven until the threshold time. Among the sites that reach the
threshold, we select
\begin{equation}
\begin{aligned}
    \mu^\star
    &= \operatorname*{arg\,min}_{\mu\in\{1,\ldots,N\}}T_\mu,\\
    t^{\prime\star}
    &= t_{\mu^\star}^{\prime\star}.
\end{aligned}
\label{eq:best-site-and-turn-on-time}
\end{equation}
If no site reaches the threshold within $t_{\max}$, no optimal site is
assigned.

Because the error contains sums of complex exponentials, the
stationarity equation generally has no closed-form root. Numerically,
we bracket its roots using the analytical derivative in
Eq.~\eqref{eq:error-delay-derivative}, compare the error at the resulting
stationary points and at both boundaries, and use a coarse time scan to
bracket the earliest threshold crossing. Root-finding then refines that
crossing and the corresponding switching time.

\section*{Acknowledgments}

\section*{Data Availability}
\textcolor{black}{The data supporting the findings of this study will be made publicly available upon submission of the manuscript. The repository link and DOI will be added once the data have been deposited.}


\end{document}